\documentclass{aa}
\usepackage{graphicx}
\usepackage{amssymb}
\begin{document}
\title{VLTI/VINCI diameter constraints on the evolutionary status of \quad $\delta$ Eri, $\xi$ Hya, $\eta$ Boo}
\author{
F. Th\'evenin\inst{1},
P. Kervella\inst{2},
B. Pichon \inst{1},
P. Morel \inst{1},
E. Di Folco\inst{3},
\and
Y. Lebreton\inst{4} 
}

\institute{ Laboratoire Cassiop\'ee, UMR 6202 CNRS, Observatoire
de la C\^ote d'Azur, BP 4229, 06304 Nice Cedex 4, France
\and
Laboratoire d'Etudes Spatiales et d'Instrumentation Astrophysique (LESIA), UMR 8109 du CNRS,
Observatoire de Paris, Section de Meudon, 5 place Jules Janssen, 92195 Meudon Cedex, France
\and
European Southern Observatory, Karl-Schwarzschild Strasse 2, D-85748 Garching,
Germany
\and 
GEPI, UMR 8111 CNRS, Observatoire de Paris, Section de Meudon, 5 place 
Jules Janssen, 92195 Meudon Cedex, France
} 
\titlerunning{$\delta$ Eri, $\xi$ Hya, $\eta$ Boo}
\authorrunning{Th\'evenin et al.}

\mail{thevenin@obs-nice.fr}
\date{Received ; Accepted }

\abstract{Using VLTI/VINCI angular diameter measurements, we constrain the
 evolutionary status of three asteroseismic targets: the stars $\delta$ Eri, $\xi$ Hya, $\eta$ Boo.
 Our predictions of the mean large frequency spacing of these stars are in agreement with 
 published observational estimations.
 Looking without success for a companion of $\delta$ Eri we doubt on its
 classification as an RS CVn star. }
\maketitle
\section{Introduction}
After two years of operation, the commissioning instrument 
VINCI of the VLTI has provided valuable
stellar diameter measurements. Among the impact of these diameters are the
studies of main sequence stars, where diameters combined with asteroseismic 
frequencies can be used to constrain evolutionary status and mass.
Several papers have been subsequently published 
(S\'egransan et al. \cite{segransan03}, Kervella et al. \cite{kervella03a}, \cite{kervella03b}, 
\cite{kervella04a} and Di Folco et al. \cite{difolco04})
  with important results on stellar fundamental parameters prior 
  to the use of the dedicated VLTI light combiner: AMBER (Petrov et al. \cite{petrov03}).
The aim of the present paper is to complete previous studies using VINCI
  to measure the diameter of three subgiant and giant stars which are among 
  selected asteroseismic targets for ground-based observations and space missions:
  $\delta$ Eri, $\xi$ Hya, $\eta$ Boo.
We perform a preliminary study of their evolutionary status by constraining their mass, 
  their helium content and their age.
One of the purpose of this paper is to show that in the future, the use of 
  stellar diameters will be a significant constraint for evolutionary models for a given input physics. 
We first detail the characteristics of each of the three stars 
  (Sect.~\ref{status}) and then we present diameter measurements (Sect.~\ref{dia})
  for each star. 
We construct evolutionary models satisfying spectro-photometric
 observable constraints and we confront asteroseismic large frequencies with measured ones.
We present these models (Sect.~\ref{modeles})  and we draw some conclusions on the
  classification and fundamental parameters of the three stars.
\section{Global characteristics of the stars}
\label{status}

The first part of table \ref{tableau} presents the observational data of the three stars. 
  The second part of this table summarizes some input parameters and output data of the models.

\def\saut{\hspace{8.4truemm}}
\begin{table*}
\caption[]{ \label{tableau} Observable characteristics of the stars and best model
  reproducing them. The subscripts ``$_{\rm ini}$'' and ``$_{\rm surf}$'' respectively refer
  to initial values and surface quantities at present day. 
  Note that the presented errors of VLTI/VINCI angular diameters are the statistical 
  ones followed by the systematical ones.
  Note also that, in any cases, $\rm D/D_\odot $ is equal to $ \rm  R/R_\odot $.
  }
\begin{tabular}{lllllll} \\
\hline
\tabularnewline[0.5\baselineskip]
                          &   $\delta$ Eri        & & $\xi$ Hya           & & $\eta$ Boo          &  \\
  $ \rm V           $     &   $  3.51 \pm 0.02  $ & & $  3.54 \pm 0.01  $ & & $  2.68 \pm 0.01  $ & \\ 
  $ \rm BC          $     &   $ -0.24 \pm 0.01  $ & & $ -0.26 \pm 0.01  $ & & $ -0.06 \pm 0.01  $ & \\
  $ \rm T_{eff} (K) $     &   $  5074 \pm 60    $ & & $  5010 \pm 100   $ & & $  6050 \pm 150   $ & \\
  $\rm  L/L_\odot   $     &   $  3.19 \pm 0.06  $ & & $  60.7 \pm 4.1   $ & & $  8.95 \pm 0.20  $ & \\
  $\rm  [Fe/H]_{surf} $   &   $  0.13 \pm 0.08  $ & & $ -0.04 \pm 0.12  $ & & $  0.24 \pm 0.07  $ & \\
  $\rm  \log g    $       &   $  3.77 \pm 0.16  $ & & $  2.93 \pm 0.30  $ & & $  3.66 \pm 0.20  $ & \\
  $\rm \theta_{LD}(mas) $ &   $ 2.394 \pm 0.014 $ & & $ 2.386 \pm 0.009 $ & & $ 2.200 \pm 0.027 $ & \\
                          & \saut $   \pm 0.025 $ & & \saut $ \pm 0.019 $ & & \saut $ \pm 0.016 $ & \\
  $\rm D/D_\odot  $       &   $ 2.33 \pm 0.03   $ & & $  10.3 \pm 0.3   $ & & $  2.68 \pm 0.05 $ & \\
  $\rm \pi (mas) $        &   $ 110.58 \pm 0.88 $ & & $ 25.23 \pm 0.83  $ & & $ 88.17 \pm 0.75  $ & \\
  $\rm  \Delta \nu_0 
       (\mu Hz)$          &   $ 43.8 \pm 0.3    $ & & $ 7.1             $ & & $ 40.47 \pm 0.05  $ & \\

\tabularnewline[0.5\baselineskip]
\hline                                                                                    
\tabularnewline[0.5\baselineskip]
            & $\delta$ Eri    & $\delta$ Eri &               $\xi$ Hya    & $\xi$ Hya &           $\eta$ Boo & $\eta$ Boo \\
              &    diffusion   & no diffusion 
                                     &  diffusion & no diffusion   & diffusion &   no diffusion\\
\tabularnewline[0.2\baselineskip]
 $ \rm  M / M_\odot $    & 1.215  &  1.215   &   2.65   & 2.65     &    1.70   &      1.70     \\
  age of
  the ZAMS 
  (Myr)                  & 20.14  &  20.06   &  2.724   &  2.719   &    12.68  &      12.67    \\
age (from
 ZAMS) (Myr)             & 6194.  &  6196.   &  509.52  &  505.34  &    2738.5 &      2355.    \\
 $ \rm Y_{ini} $         & 0.28   &  0.28    &  0.275   &  0.275   &    0.260  &      0.260    \\
 $\rm  [Z/X]_{ini} $     & 0.148  &  0.148   &  0.00    &  0.00   &    0.367  &      0.367    \\
 $ \rm T_{eff} (K) $     & 5055.  &  5066.   &  5037.   &  5034.   &    6050.  &      6090.    \\
 $\rm  L/L_\odot  $      & 3.176  &  3.230   &  61.23   &  61.0    &    8.944  &      8.978    \\
 $ \rm  R/R_\odot  $     & 2.328  &  2.337   &  10.30   &  10.30   &    2.728  &      2.697    \\
 $\rm  \log g   $        & 3.788  &  3.785   &  2.835   &  2.832   &    3.796  &      3.806    \\
 $\rm  Y_{surf}  $       & 0.266  &  0.28    &  0.274   &  0.275   &    0.228  &      0.260    \\
 $\rm  [Z/X]_{surf}$     & 0.123  &  0.148   &  0.00    &  0.00    &    0.303  &      0.367    \\
 $\rm M_{CZ} (M_{\star})$& 0.729  &  0.727   &  0.608   &  0.596   &    0.9994 &      0.9994   \\ 
 $\rm R_{CZ} (R_{\star})$& 0.475  &  0.475   &  0.422   &  0.417   &    0.8388 &      0.8505   \\ 
  $\rm  \Delta 
  \nu_0 (\mu Hz)$     & 45.27  &  44.91   &  7.23    &  7.28    &      41.91  &      42.47 \\
\tabularnewline[0.5\baselineskip]
\hline
\end{tabular}
\end{table*}

\subsection{$\delta$ Eri}
$\delta$ Eri (HD 23249, HR 1136, HIP 17378) has been thoroughly studied by
  photometry and spectroscopy and is classified as a K0\,IV star (Keenan \& Pitts \cite{keenan80}). 
It belongs to the group of the nearest stars with an accurate Hipparcos
  parallax of $ \rm 110.58 \pm 0.88 \, mas$ (Perryman et al. \cite{perryman97}).
The star has been classified as weakly active and X-ray soft source 
  (Huensch et al. \cite{huensch99}) after a long time of search for its activity.
Wilson \& Bappu (\cite{wilson57}) concluded that a possible detection of
  emission in the lines H\&K is "{\sl exceedingly weak}" - so weak that it is 
  questionable.
Finally, it took more than 20 years to really detect its activity 
  with Copernicus revealing a weak emission in MgII (Weiler \& Oegerle \cite{weiler79}).
  Fisher et al. (\cite{fisher83}) tried to detect a periodic variation in the photometric 
  data and concluded that, if it exists, the amplitude is below $ \pm 0.02 $ magnitude 
  with a period of 10 days.
They suggested that $\delta$ Eri could be classified as a RS CVn star.
A RS CVn is defined as a F-G binary star having a period shorter than 14 days, with
  a chromospheric activity and with a period of rotation synchronized with its orbital
  period (Linsky \cite{linsky84}) then giving to the star a high rotational
  velocity inducing a strong activity.
All of this is in contrast with the very small activity
  detected for $\delta$ Eri making doubtful its classification as a RS CVn star.
  $\delta$ Eri having a projected rotational velocity of $\rm v \sin i = 1.0 \, km \, s^{-1}$ 
  (de Meideros \& Mayor \cite{medeir99}) the hypothetical RS CVn classification forces 
  us to conclude that the binary is seen pole-on therefore explaining the lack of photometric
  variations and also of any variation of the radial velocity (Santos et al. \cite{santos04}).
In attempting to reveal the presence of a close companion around  $\delta$~Eri, 
  we set several VLTI/VINCI observations at different baselines (see Sect.~\ref{dia}).

We estimate its bolometric luminosity to 
  $\rm L_{\star}/L_{\odot} = 3.19 \pm 0.06 $ using Alonso et al. (\cite{alonso99})
  empirical bolometric corrections (BC, $\rm BC = -0.24 \pm 0.01 $ for giants,
  this latter is the dominant source of uncertainty on luminosity).
We adopt Santos et al. (\cite{santos04}) values  for the effective temperature 
  $ \rm T_{eff} = 5074. \pm 60. \, K $, 
  logarithmic surface gravity $\log g = 3.77 \pm 0.16 $ and surface 
  iron abundance $ \rm [Fe/H] = 0.13 \pm 0.03 $.
These parameters are different from -- but within the error bars of -- the 
  parameters proposed by Pijpers (\cite{pijpers03}) for this star, except the metallicity 
  which is 0.24 dex higher.
Bouchy \& Carrier (\cite{bouchy03}) have measured a mean large frequency
  spacing of $ \rm 43.8 \mu Hz $ that we will try to reproduce with our model.
We recall that the large frequency spacing is defined as the difference between
  frequencies of modes with consecutive radial order $n$~:
  $\Delta \nu_l (n) = \nu_{n,l}- \nu_{n-1,l} $.   
  In the high frequency range, i.e. large radial orders 
  $\Delta \nu_l (n) $ is almost constant with a mean value 
  strongly related to the square root of the mean density of the star.
  To obtain the mean large frequency separation, we average over $ l = 0 - 2 $.
\subsection{$\xi$ Hya}
$\xi$ Hya (HD 100407, HR 4450, HIP 56343)
  is a giant star (G7\,III) which has been considered by Eggen (\cite{eggen77}) 
  as a spurious member of the Hyades group because it departs slightly from
  the regression line of giant stars in the colour diagrams 
  (b-y,R-I) and ($\rm M_1$,R-I) of that stellar group. 

Its Hipparcos parallax is $ \rm 25.23 \pm 0.83 \, mas $. 
We estimate its bolometric luminosity to $\rm L_{\star}/L_{\odot} = 60.7 \pm 4.1 $ 
  using BC ($\rm BC = -0.26 \pm 0.01 $) from Alonso et al. (\cite{alonso99}).
We adopt the spectroscopic parameters derived by Mc William (\cite{mcwilliam}): 
   effective temperature $ \rm T_{eff} = 5010. \pm 100. \, K $, 
   $ \log g = 2.93 \pm 0.30 $ and $ \rm [Fe/H] = -0.04 \pm 0.12 $.
These parameters are different from -- but within the error bars of -- the 
  parameters adopted by Frandsen et al. (\cite{frandsen02}) for this star.
The star belongs to the HR diagram at the lowest
   part of the giant branch corresponding to an evolved star with a mass around 
   $ \rm 3 M_{\odot} $. 
Using a set of CORALIE spectra, Frandsen et al. (\cite{frandsen02}) 
  detected solar-like oscillations suggesting radial modes with the largest amplitudes 
  almost equidistant around $ \rm 7.1 \mu Hz $. 
That important detection opens the possibility to better
  constrain the model of that star for which the mass is not well-known. 
\subsection{$\eta$ Boo}
$\eta$ Boo (HD 121370, HR 5235, HIP 67927) is a subgiant (G0\,IV) spectroscopic binary (SB1)
   studied recently by Di Mauro et al. (\cite{dimauro03}, \cite{dimauro04}) and Guenther (\cite{guenther04}). 
Its Hipparcos parallax is $ \rm 88.17 \pm 0.75 \, mas $. 
Having large overabundances of Si, Na, S, Ni and Fe, it has been
  considered as super-metal-rich by Feltzing \& Gonzales (\cite{feltzing01}).
We adopt here a luminosity $\rm L_{\star}/L_{\odot} = 8.95 \pm 0.20 $ using BC 
  ($\rm BC = -0.06 \pm 0.01 $, this latter is the dominant source of 
  uncertainty on luminosity) from Vandenberg and Clem (\cite{vandenberg03}) for this subgiant,
  an effective temperature $ \rm T_{eff} = 6050. \pm 150. \, K $
  representing the average of five effective temperature determinations in the [Fe/H]
  catalogue of Cayrel de Strobel et al. (\cite{giusa01}) and the spectroscopic 
  $ \rm \log g = 3.66 \pm 0.20 $ and $ \rm [Fe/H] = 0.24 \pm 0.07 $ from 
  Feltzing \& Gonzales (\cite{feltzing01}).
These parameters are different from -- but within the error bars of -- the 
  parameters adopted by Di Mauro et al. (\cite{dimauro03}, \cite{dimauro04}) for this star.
Asteroseismic observations of $ \delta $ Eri have been reported
  by Carrier et al. (\cite{carrier05}) with $ \rm \Delta \nu_0 = 39.9 \pm 0.1 \, \mu Hz $  
  and by Kjeldsen et al. (\cite{kjeldsen03}) with $ \rm \Delta \nu_0 = 40.47 \pm 0.05 \, \mu Hz $.
\section{Diameter interferometric measurements}
\label{dia}
\subsection{VINCI and the VLTI}
 
The European Southern Observatory's Very Large Telescope Interferometer
  (Glindemann et al. \cite{glindemann}) is operated  on top of the Cerro Paranal, in Northern Chile since
  March 2001. 
For the observations reported in this work, the light coming from two telescopes (two 0.35m test siderostats or
  VLT/UT1-UT3) was combined coherently in VINCI, the VLT Interferometer Commissioning Instrument
  (Kervella et al. \cite{kervella00}).  
We used a regular $K$ band filter ($ \rm \lambda = 2.0 - 2.4 \ \mu m$) for these observations.

\subsection{Data reduction}
We used an improved version of the standard VINCI data 
  reduction pipeline (Kervella, S\'egransan \& Coud\'e du Foresto~\cite{kervella04b}), 
  whose general principle
  is based on the original FLUOR algorithm (Coud\'e du Foresto et al. \cite{cdf97}).
The two calibrated output interferograms are subtracted to remove residual
  photometric fluctuations. 
Instead of the classical Fourier analysis, we implemented a time-frequency 
  analysis (S\'egransan et al. \cite{s99}) based on a continuous wavelet transform.


%


The atmospheric piston effect between the two telescopes corrupts the amplitude 
  and the shape of the fringe peak in the wavelet power spectrum. 
As described in Kervella et al. (\cite{kervella04b}), the properties of the fringe
  peaks in the time and frequency domains are monitored automatically, in order 
  to reject from the processing the interferograms that are strongly affected by 
  the atmospheric piston.
This selection reduces the statistical dispersion of the squared coherence 
  factors ($\mu^2$) measurement, and avoids biases from corrupted interferograms.
The final $\mu^2$ values are derived by integrating the average wavelet power 
  spectral density (PSD) of the interferograms at the position and frequency of 
  the fringes.
The residual photon and detector noise backgrounds are removed using a linear 
  least squares fit of the PSD at high and low frequency. 
The statistical error bars on $\mu^2$ are computed from the series of $\mu^2$ 
  values obtained on each target star (typically a few hundreds interferograms) 
  using the bootstrapping technique.
\subsection{Measured visibilities and angular diameters}
\label{vis_values}
\begin{figure}[t]
\centering
\includegraphics[width=8.5cm]{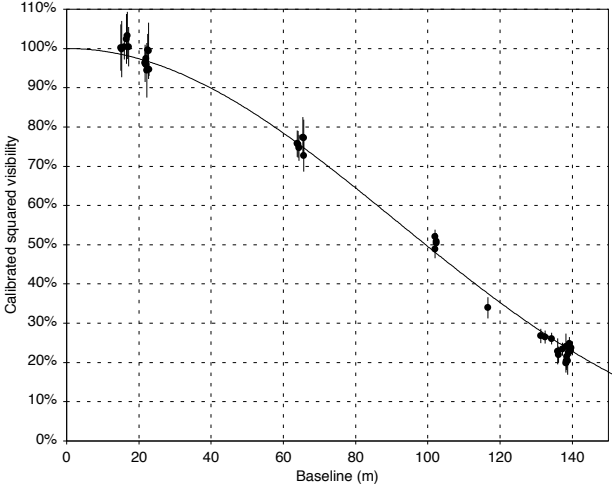}
\caption{ \label{fig:del_eri_vis} Squared visibility measurements obtained on  $ \rm \delta \, Eri$.
The solid line is a limb darkened disk model with $ \rm \theta_{LD} = 2.394 \pm 0.014 \pm 0.025 \, mas$ 
  (statistical and systematic errors).}
\end{figure}

\begin{figure}[t]
\centering
\includegraphics[width=8.5cm]{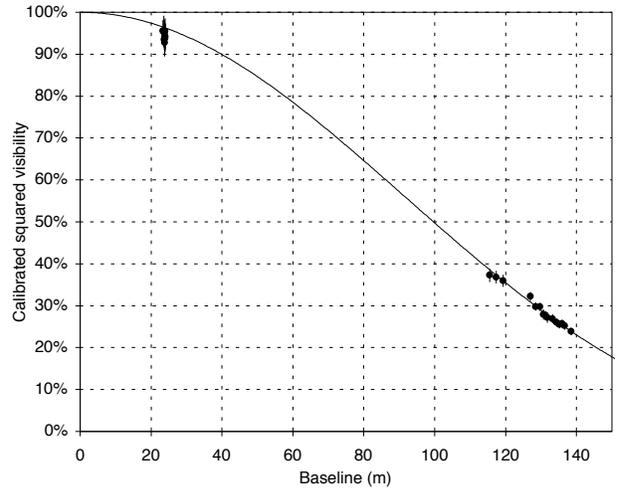}
\caption{ \label{fig:ksi_hya_vis} Squared visibility measurements obtained on $ \rm \xi \,Hya$.
The solid line is a limb darkened disk model with $ \rm \theta_{LD} = 2.386 \pm 0.009 \pm 0.019 \, mas$ 
  (statistical and systematic errors).}
\end{figure}

\begin{figure}[t]
\centering
\includegraphics[width=8.5cm]{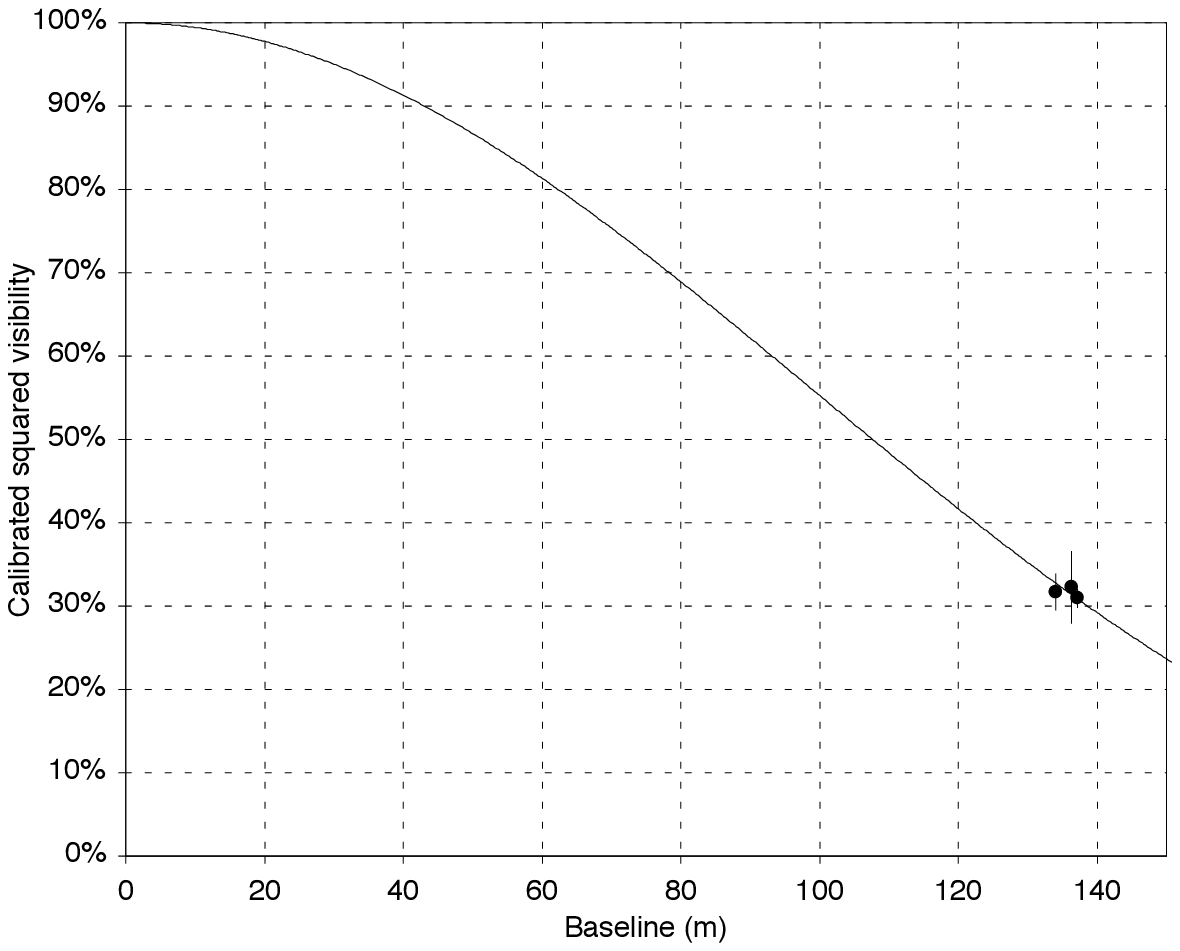}
\caption{ \label{fig:eta_boo_vis} Squared visibility measurements obtained on $\rm \eta \, Boo$.
The solid line is a limb darkened disk model with $ \rm \theta_{LD} = 2.200 \pm 0.027 \pm 0.016 \, mas$ 
  (statistical and systematic errors).}
\end{figure}

The visibility values obtained on $\delta$\,Eri, $\xi$\,Hya and $\eta$\,Boo are listed in
Tables~\ref{table_vis_deleri1} to \ref{table_vis_etaboo}, and plotted on Figures~\ref{fig:del_eri_vis}
to \ref{fig:eta_boo_vis}. 

The calibration of the visibilities obtained on $\delta$\,Eri and $\eta$\,Boo was done using
well-known calibrator stars that were selected in the Cohen et al. (\cite{cohen99}) catalogue.
The uniform disk (UD) angular diameter of these stars was converted into a limb darkened value
and then to a $K$ band uniform disk angular diameter using the recent non-linear law
coefficients taken from Claret et al. (\cite{claret00}). As demonstrated by
Bord\'e et al. (\cite{borde}), the star diameters in this list have been measured very homogeneously
to a relative precision of approximately 1\%. 

\begin{table*}
\caption{ \label{table_vis_deleri1} $\delta$\,Eri squared visibilities.}
\begin{tabular}{ccccccl}
\hline
Julian Date & Stations & $N$ & $B$ (m) & Az. ($\deg$) & $V^2 \pm {\rm stat} \pm {\rm syst}$ & Calibrator\\
\hline
2452682.528&B3-D1& $74$ & $22.638$ & $14.95$ & $0.9941\pm0.0712\pm0.0014$ &$\delta$ Lep\\
2452682.541&B3-D1& $460$ & $21.963$ & $14.63$ & $0.9740\pm0.0140\pm0.0014$ &$\delta$ Lep\\
2452682.545&B3-D1& $281$ & $21.735$ & $14.55$ & $0.9639\pm0.0264\pm0.0014$ &$\delta$ Lep\\
2452682.607&B3-D1& $140$ & $16.514$ & $14.78$ & $1.0242\pm0.0632\pm0.0014$ &$\delta$ Lep\\
2452682.612&B3-D1& $340$ & $15.954$ & $14.99$ & $1.0045\pm0.0321\pm0.0014$ &$\delta$ Lep\\
2452682.618&B3-D1& $133$ & $15.285$ & $15.27$ & $0.9987\pm0.0715\pm0.0013$ &$\delta$ Lep\\
2452671.562&B3-D1& $233$ & $22.437$ & $14.84$ & $0.9960\pm0.0409\pm0.0031$ &$\delta$ Lep\\
2452671.567&B3-D1& $95$ & $22.164$ & $14.71$ & $0.9442\pm0.0697\pm0.0029$ &$\delta$ Lep\\
2452671.574&B3-D1& $210$ & $21.749$ & $14.55$ & $0.9623\pm0.0474\pm0.0030$ &$\delta$ Lep\\
2452671.631&B3-D1& $397$ & $17.152$ & $14.59$ & $1.0042\pm0.0501\pm0.0014$ &$\delta$ Lep\\
2452671.635&B3-D1& $206$ & $16.756$ & $14.71$ & $1.0331\pm0.0604\pm0.0014$ &$\delta$ Lep\\
2452671.651&B3-D1& $237$ & $14.947$ & $15.44$ & $1.0023\pm0.0588\pm0.0014$ &$\delta$ Lep\\
2452672.553&B3-D1& $401$ & $22.756$ & $15.02$ & $0.9465\pm0.0164\pm0.0014$ &$\delta$ Lep\\
2452672.567&B3-D1& $426$ & $22.013$ & $14.65$ & $0.9585\pm0.0153\pm0.0014$ &$\delta$ Lep\\
2452672.603&B3-D1& $379$ & $19.478$ & $14.26$ & $0.9911\pm0.0235\pm0.0014$ &$\delta$ Lep\\
2452672.607&B3-D1& $237$ & $19.086$ & $14.28$ & $1.0134\pm0.0540\pm0.0015$ &$\delta$ Lep\\
2452673.567&B3-D1& $236$ & $21.898$ & $14.60$ & $0.9780\pm0.0322\pm0.0014$ &$\delta$ Lep\\
2452673.579&B3-D1& $264$ & $21.130$ & $14.39$ & $0.9940\pm0.0264\pm0.0015$ &$\delta$ Lep\\
2452673.609&B3-D1& $441$ & $18.693$ & $14.31$ & $1.0197\pm0.0253\pm0.0015$ &$\delta$ Lep\\
2452674.527&B3-D1& $262$ & $23.527$ & $15.78$ & $0.9718\pm0.0294\pm0.0014$ &$\delta$ Lep\\
2452674.557&B3-D1& $415$ & $22.253$ & $14.75$ & $0.9757\pm0.0241\pm0.0015$ &$\delta$ Lep\\
2452674.562&B3-D1& $405$ & $22.003$ & $14.64$ & $0.9833\pm0.0249\pm0.0015$ &$\delta$ Lep\\
2452674.566&B3-D1& $314$ & $21.756$ & $14.55$ & $0.9778\pm0.0281\pm0.0015$ &$\delta$ Lep\\
2452675.547&B3-D1& $432$ & $22.640$ & $14.95$ & $0.9731\pm0.0213\pm0.0014$ &$\delta$ Lep\\
2452676.557&B3-D1& $383$ & $21.997$ & $14.64$ & $0.9674\pm0.0203\pm0.0014$ &$\delta$ Lep\\
2452676.561&B3-D1& $402$ & $21.734$ & $14.55$ & $0.9813\pm0.0201\pm0.0015$ &$\delta$ Lep\\
2452676.565&B3-D1& $259$ & $21.474$ & $14.47$ & $0.9678\pm0.0338\pm0.0014$ &$\delta$ Lep\\
2452676.590&B3-D1& $447$ & $19.612$ & $14.26$ & $0.9883\pm0.0227\pm0.0014$ &$\delta$ Lep\\
2452676.602&B3-D1& $328$ & $18.603$ & $14.32$ & $0.9453\pm0.0318\pm0.0013$ &$\delta$ Lep\\
2452677.543&B3-D1& $480$ & $22.582$ & $14.92$ & $0.9651\pm0.0283\pm0.0014$ &$\delta$ Lep\\
2452677.547&B3-D1& $445$ & $22.366$ & $14.80$ & $0.9695\pm0.0294\pm0.0014$ &$\delta$ Lep\\
2452677.551&B3-D1& $256$ & $22.137$ & $14.70$ & $0.9283\pm0.0407\pm0.0013$ &$\delta$ Lep\\
2452677.587&B3-D1& $267$ & $19.633$ & $14.26$ & $1.0093\pm0.0407\pm0.0015$ &$\delta$ Lep\\
2452677.598&B3-D1& $381$ & $18.695$ & $14.31$ & $1.0013\pm0.0384\pm0.0015$ &$\delta$ Lep\\
2452677.603&B3-D1& $287$ & $18.286$ & $14.36$ & $1.0432\pm0.0455\pm0.0015$ &$\delta$ Lep\\
2452678.537&B3-D1& $230$ & $22.746$ & $15.02$ & $1.0024\pm0.0382\pm0.0014$ &$\delta$ Lep\\
2452678.548&B3-D1& $121$ & $22.186$ & $14.72$ & $0.9746\pm0.0520\pm0.0014$ &$\delta$ Lep\\
2452678.559&B3-D1& $168$ & $21.531$ & $14.49$ & $0.9900\pm0.0492\pm0.0014$ &$\delta$ Lep\\
2452678.584&B3-D1& $422$ & $19.649$ & $14.26$ & $1.0167\pm0.0354\pm0.0011$ &$\delta$ Lep\\
2452678.593&B3-D1& $150$ & $18.893$ & $14.29$ & $1.0966\pm0.0618\pm0.0012$ &$\delta$ Lep\\
2452679.561&B3-D1& $402$ & $21.184$ & $14.40$ & $0.9800\pm0.0353\pm0.0014$ &$\delta$ Lep\\
2452679.566&B3-D1& $278$ & $20.892$ & $14.35$ & $1.0211\pm0.0435\pm0.0015$ &$\delta$ Lep\\
2452683.578&B3-D1& $374$ & $19.065$ & $14.28$ & $0.9596\pm0.0152\pm0.0012$ &$\delta$ Lep\\
2452683.582&B3-D1& $449$ & $18.708$ & $14.31$ & $0.9900\pm0.0147\pm0.0013$ &$\delta$ Lep\\
2452683.586&B3-D1& $283$ & $18.316$ & $14.36$ & $0.9378\pm0.0232\pm0.0012$ &$\delta$ Lep\\
2452683.593&B3-D1& $269$ & $17.654$ & $14.48$ & $0.9915\pm0.0274\pm0.0013$ &$\delta$ Lep\\
2452683.598&B3-D1& $250$ & $17.167$ & $14.59$ & $0.9693\pm0.0290\pm0.0012$ &$\delta$ Lep\\
2452683.602&B3-D1& $261$ & $16.783$ & $14.70$ & $0.9154\pm0.0274\pm0.0012$ &$\delta$ Lep\\
2452684.516&B3-D1& $296$ & $22.937$ & $15.15$ & $0.9431\pm0.0287\pm0.0014$ &$\delta$ Lep\\
2452684.527&B3-D1& $400$ & $22.396$ & $14.82$ & $0.9473\pm0.0220\pm0.0014$ &$\delta$ Lep\\
2452684.562&B3-D1& $439$ & $20.148$ & $14.27$ & $0.9859\pm0.0225\pm0.0013$ &$\delta$ Lep\\
2452684.579&B3-D1& $415$ & $18.747$ & $14.30$ & $0.9882\pm0.0232\pm0.0013$ &$\delta$ Lep\\
2452685.587&B3-D1& $206$ & $17.669$ & $14.47$ & $1.0318\pm0.0277\pm0.0013$ &$\delta$ Lep\\
\hline
\end{tabular}
\end{table*}

\begin{table*}
\caption{ \label{table_vis_deleri2} $\delta$\,Eri squared visibilities 
(continued from Table~\ref{table_vis_deleri1}).}
\begin{tabular}{ccccccl}
\hline
Julian Date & Stations & $N$ & $B$ (m) & Az. ($\deg$) & $V^2 \pm {\rm stat} \pm {\rm syst}$ & Calibrator\\
\hline
2452524.854&E0-G1& $350$ & $65.689$ & $307.62$ & $0.7271\pm0.0400\pm0.0054$ &70\,Aql, 31\,Ori\\
2452524.858&E0-G1& $336$ & $65.583$ & $307.23$ & $0.7720\pm0.0464\pm0.0057$ &70\,Aql, 31\,Ori\\
2452524.863&E0-G1& $239$ & $65.450$ & $306.79$ & $0.7729\pm0.0521\pm0.0057$ &70\,Aql, 31\,Ori\\
2452524.890&E0-G1& $452$ & $64.342$ & $303.74$ & $0.7467\pm0.0329\pm0.0055$ &70\,Aql, 31\,Ori\\
2452524.895&E0-G1& $456$ & $64.115$ & $303.16$ & $0.7561\pm0.0336\pm0.0056$ &70\,Aql, 31\,Ori\\
2452524.899&E0-G1& $452$ & $63.877$ & $302.56$ & $0.7579\pm0.0332\pm0.0056$ &70\,Aql, 31\,Ori\\
2452555.889&B3-M0& $312$ & $132.444$ & $27.46$ & $0.2742\pm0.0150\pm0.0055$ &$\delta$ Phe\\
2452555.893&B3-M0& $275$ & $131.275$ & $27.44$ & $0.2769\pm0.0168\pm0.0056$ &$\delta$ Phe\\
2452556.810&B3-M0& $200$ & $139.144$ & $30.60$ & $0.2477\pm0.0152\pm0.0067$ &$\delta$ Phe\\
2452556.817&B3-M0& $395$ & $139.500$ & $30.10$ & $0.2294\pm0.0113\pm0.0062$ &$\delta$ Phe\\
2452556.822&B3-M0& $373$ & $139.635$ & $29.80$ & $0.2370\pm0.0117\pm0.0064$ &$\delta$ Phe\\
2452564.830&B3-M0& $146$ & $138.416$ & $28.23$ & $0.2047\pm0.0228\pm0.0019$ &HR\,8685\\
2452567.762&B3-M0& $236$ & $137.272$ & $32.21$ & $0.2245\pm0.0153\pm0.0044$ &HR\,8685\\
2452577.789&B3-M0& $173$ & $138.926$ & $28.46$ & $0.2248\pm0.0314\pm0.0070$ &45\,Eri, HR\,2549\\
2452577.794&B3-M0& $187$ & $138.426$ & $28.23$ & $0.2156\pm0.0289\pm0.0067$ &45\,Eri, HR\,2549\\
2452213.776&UT1-UT3& $73$ & $101.996$ & $232.98$ & $0.4883\pm0.0203\pm0.0102$ &$\chi$\,Phe\\
2452213.777&UT1-UT3& $332$ & $102.056$ & $232.83$ & $0.5207\pm0.0138\pm0.0109$ &$\chi$\,Phe\\
2452213.791&UT1-UT3& $69$ & $102.374$ & $231.76$ & $0.5089\pm0.0172\pm0.0106$ &$\chi$\,Phe\\
2452213.793&UT1-UT3& $312$ & $102.394$ & $231.65$ & $0.5044\pm0.0150\pm0.0105$ &$\chi$\,Phe\\
2452578.723&B3-M0& $269$ & $135.965$ & $33.09$ & $0.2393\pm0.0257\pm0.0063$ &$\tau$\,Cet\\
2452578.740&B3-M0& $169$ & $138.202$ & $31.51$ & $0.2520\pm0.0246\pm0.0066$ &$\tau$\,Cet\\
2452578.745&B3-M0& $74$ & $138.752$ & $31.02$ & $0.2133\pm0.0307\pm0.0056$ &$\tau$\,Cet\\
2452585.799&B3-M0& $298$ & $134.322$ & $27.55$ & $0.2608\pm0.0134\pm0.0071$ &$\tau$\,Cet\\
2452601.810&B3-M0& $206$ & $116.676$ & $28.41$ & $0.3674\pm0.0290\pm0.0082$ &$\tau$\,Cet\\
2452602.728&B3-M0& $123$ & $138.193$ & $28.15$ & $0.2183\pm0.0241\pm0.0056$ &$\tau$\,Cet\\
2452602.742&B3-M0& $396$ & $136.193$ & $27.73$ & $0.2412\pm0.0174\pm0.0062$ &$\tau$\,Cet\\
\hline
\end{tabular}
\end{table*}

\begin{table*}
\caption{ \label{table_vis_xihya} $\xi$\,Hya squared visibilities.}
\begin{tabular}{ccccccl}
\hline
Julian Date & Stations & $N$ & $B$ (m) & Az. ($\deg$) & $V^2 \pm {\rm stat} \pm {\rm syst}$ & Calibrators\\
\hline
2452681.743&B3-D1& $333$ & $23.650$ & $27.39$ & $0.9539\pm0.0376\pm0.0008$ &$\alpha$ Crt\\
2452681.747&B3-D1& $460$ & $23.727$ & $26.48$ & $0.9520\pm0.0305\pm0.0008$ &$\alpha$ Crt\\
2452681.751&B3-D1& $343$ & $23.801$ & $25.51$ & $0.9281\pm0.0334\pm0.0007$ &$\alpha$ Crt\\
2452681.777&B3-D1& $452$ & $23.995$ & $20.50$ & $0.9555\pm0.0304\pm0.0008$ &$\alpha$ Crt\\
2452681.781&B3-D1& $332$ & $23.989$ & $19.60$ & $0.9383\pm0.0337\pm0.0008$ &$\alpha$ Crt\\
2452681.785&B3-D1& $427$ & $23.975$ & $18.89$ & $0.9424\pm0.0305\pm0.0008$ &$\alpha$ Crt\\
2452682.729&B3-D1& $354$ & $23.407$ & $29.82$ & $0.9560\pm0.0251\pm0.0009$ &$\alpha$ Crt\\
2452682.752&B3-D1& $295$ & $23.846$ & $24.85$ & $0.9519\pm0.0280\pm0.0009$ &$\alpha$ Crt\\
2452682.792&B3-D1& $297$ & $23.904$ & $17.19$ & $0.9420\pm0.0317\pm0.0007$ &$\alpha$ Crt\\
2452682.801&B3-D1& $403$ & $23.773$ & $15.47$ & $0.9351\pm0.0237\pm0.0007$ &$\alpha$ Crt\\
2452760.583&B3-M0& $350$ & $138.521$ & $60.37$ & $0.2383\pm0.0058\pm0.0069$ &$\alpha$ Crt\\
2452760.600&B3-M0& $343$ & $136.690$ & $63.11$ & $0.2520\pm0.0061\pm0.0073$ &$\alpha$ Crt\\
2452760.605&B3-M0& $391$ & $135.918$ & $63.96$ & $0.2568\pm0.0059\pm0.0075$ &$\alpha$ Crt\\
2452760.635&B3-M0& $433$ & $129.762$ & $68.49$ & $0.2971\pm0.0058\pm0.0077$ &$\alpha$ Crt\\
2452760.640&B3-M0& $388$ & $128.458$ & $69.20$ & $0.2978\pm0.0061\pm0.0077$ &$\alpha$ Crt\\
2452760.645&B3-M0& $284$ & $127.037$ & $69.92$ & $0.3221\pm0.0071\pm0.0084$ &$\alpha$ Crt\\
2452761.624&B3-M0& $429$ & $131.833$ & $67.24$ & $0.2714\pm0.0063\pm0.0097$ &51 Hya\\
2452761.628&B3-M0& $303$ & $130.716$ & $67.94$ & $0.2787\pm0.0077\pm0.0100$ &51 Hya\\
2452761.665&B3-M0& $421$ & $119.296$ & $73.16$ & $0.3592\pm0.0063\pm0.0131$ &51 Hya\\
2452761.671&B3-M0& $402$ & $117.300$ & $73.87$ & $0.3681\pm0.0067\pm0.0135$ &51 Hya\\
2452761.675&B3-M0& $340$ & $115.485$ & $74.49$ & $0.3727\pm0.0087\pm0.0136$ &51 Hya\\
2452762.604&B3-M0& $470$ & $135.192$ & $64.66$ & $0.2554\pm0.0021\pm0.0092$ &51 Hya\\
2452762.609&B3-M0& $454$ & $134.296$ & $65.44$ & $0.2600\pm0.0022\pm0.0094$ &51 Hya\\
2452762.614&B3-M0& $386$ & $133.310$ & $66.21$ & $0.2689\pm0.0049\pm0.0097$ &51 Hya\\
2452762.623&B3-M0& $441$ & $131.274$ & $67.59$ & $0.2771\pm0.0027\pm0.0100$ &51 Hya\\
\hline
\end{tabular}
\end{table*}
\begin{table*}
\caption{ \label{table_vis_etaboo} $\eta$\,Boo squared visibilities.}
\begin{tabular}{ccccccl}
\hline
Julian Date & Stations & $N$ & $B$ (m) & Az. ($\deg$) & $V^2 \pm {\rm stat} \pm {\rm syst}$ & Calibrators\\
\hline
2452760.684&B3-M0& $131$ & $134.046$ & $64.22$ & $0.3167\pm0.0187\pm0.0119$ &$\alpha$ Crt\\
2452760.696&B3-M0& $50$ & $136.318$ & $63.31$ & $0.3227\pm0.0415\pm0.0121$ &$\alpha$ Crt\\
2452763.693&B3-M0& $187$ & $137.132$ & $62.88$ & $0.3095\pm0.0092\pm0.0064$ &$\mu$ Vir\\
\hline
\end{tabular}
\end{table*}

The VINCI instrument has no spectral dispersion and its bandpass
   corresponds to the $K$ band filter (2-2.4 $\mu$m). 
It is thus important to compute the precise effective wavelength 
   of the instrument in order to determine the angular resolution 
   at which we are observing the targets. 
The effective wavelength differs from the filter mean wavelength 
   because of the detector quantum efficiency curve, the fiber 
   beam combiner transmission and the object spectrum.
It is only weakly variable as a function of the spectral type anyway.

To derive the effective wavelength of our observations, we computed
   a model taking into account the star spectrum and the VLTI transmission.
The instrumental transmission of VINCI and the VLTI was first modeled
   taking into account all known effects and then calibrated based on several 
   bright reference stars observations with the UTs 
   (see Kervella et al. \cite{kervella03b} for details).

Taking the weighted average wavelength of this model spectrum gives an effective 
   wavelength of $\rm \lambda_{eff} = 2.178 \pm 0.003 \, \mu m $ 
   for $\delta$\,Eri, $\xi$\,Hya and $\eta$\,Boo. 
The visibility fits were computed taking into account the limb darkening 
   of the stellar disk of each stars. 
We used power law intensity profiles derived from the limb darkening 
   models of Claret (\cite{claret00}) in the $K$ band.

The resulting limb darkened diameters for the three program stars 
   are given in Table \ref{tableau}.
The statistical error bars were computed from the statistical 
  dispersion of the series of $\mu^2$ values obtained on each stars 
  (typically a few hundreds), using the bootstrapping technique. 
The systematic error bars come from the uncertainties on the angular 
  diameters of the calibrators that were used for the observation. 
They impact the precision of the interferometric transfer function 
  measurement, and thus affect the final visibility value. 
Naturally, these calibration error bars do not get smaller when the 
  number of observation increases, as the statistical errors do. 
The detailed methods and hypothesis used to compute these error bars 
  are given in Kervella et al. (\cite{kervella04b}).

\subsection{Search for a companion to $\delta$\,Eri}
\begin{figure}[t]
\centering
\includegraphics[width=8.5cm]{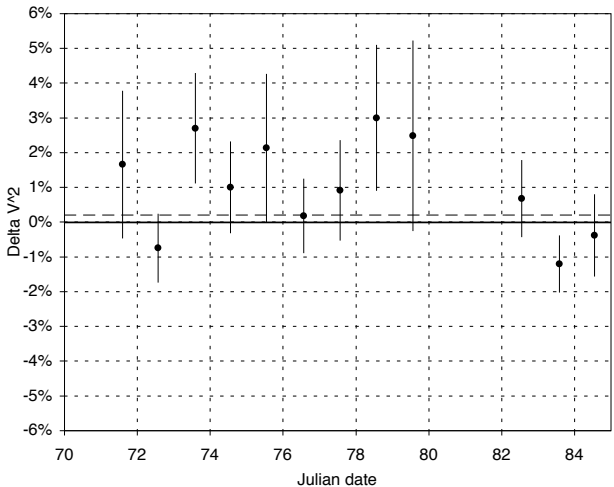}
\caption{ \label{fig:residual_del_eri} Observed deviation of the squared visibilities of $ \rm \delta \, Eri$
  (B3-D1 baseline only) with respect to the visibility model of a $ \rm \theta_{UD} = 2.394 \, mas $ 
  uniform disk model. The dashed line represents the average deviation over all observations (0.21\%).}
\end{figure}

$\delta$\,Eri is classified as an RS\,CVn variable (Kholopov et al. \cite{gcvs}), 
  and has shown a small amplitude photometric variability ($m_V$ = 3.51 to 3.56). 
  Fisher et al. (\cite{fisher83}) have also reported photometric variations with 
  an amplitude $\Delta m_V = 0.02$ over a period of 10 days.
This small amplitude and the apparent absence of periodical radial velocity 
  modulation lead these authors to propose that $\delta$\,Eri is a close binary 
  star seen nearly pole on ($i \le 5$ deg).
Following this idea, we can suggest three hypotheses to explain the observed photometric variations:
\begin{enumerate}
\item The main star is ellipsoidal. This would result in a modulation of its projected surface along the
line of sight during its rotation. This deformation would be caused by the close gravitational interaction
of the main star with the unseen companion. 
\item The companion creates a hot spot on the hemisphere of the main star that is facing it. It is
changing in apparent surface when the system rotates, probably synchronously.
\item The pole of the main component holds a dark spot that is changing in apparent surface during the
rotation of the star.
\end{enumerate}
The period of the photometric variations, if attributed to the presence of an orbiting companion,
  allows to deduce the distance between the two components through the third Kepler's law.
At the distance of $\delta$\,Eri, this corresponds to an angular separation of  approximately
  9 mas, easily resolvable using moderately long baselines of the VLTI.
Using the B3-D1 stations of the VLTI, we have taken advantage of the fact that the azimuth
  of the projected baseline is almost constant for observations of $\delta$\,Eri to monitor the
  evolution of its visibility over a period of 13 nights. 
The projected length is also very well suited to the expected separation.
Our interferometric data (Fig. \ref{fig:residual_del_eri}) does not show any systematic
  deviation from the uniform disk model fit obtained using the longer
  baselines, at a level of $0.2 \pm 0.3$\%, consistent with zero. 
From these measurements, we conclude that no companion is detected at a level of about $\pm 2$\% of the luminosity
of the primary star. This result is consistent with the fact that $\delta$\,Eri does not deviate significantly from
the surface-brightness relations determined by Kervella et al. (\cite{kervella04c}).

\section{Models and results}
\label{modeles}
In order to draw a rapid estimate of the improvements brought by the new 
  interferometric constraints on the radius on the determination of the 
  mass and age of the three stars, we have calculated evolutionary stellar 
  models that we compare to observations. 
In these models we have adopted a given set of standard input physics and 
  the observational parameters described in Section \ref{status} and Table \ref{tableau}. 
We do not intend to examine in details the effects on the uncertainties
  in the details of the models (envelope, convection, overshooting or other extra 
  mixing) on the results presented here.

The parameters used to construct our CESAM (Morel \cite{morel97}) evolutionary models 
   are summarized in Table~\ref{tableau}.
The convection is described by Canuto \& Mazitelli's theory (\cite{canuto91}, \cite{canuto92}) 
   and the atmospheres are restored on the basis of Kurucz's atlas models (\cite{kurucz92}). 
The other input physics are identical to those adopted for the star Procyon 
  (see Kervella et al. \cite{kervella04a}).
The adopted metallicity Z/X, which is an input 
   parameter for the evolutionary computations, is given by the iron abundance 
   measured in the atmosphere with the help of the following approximation: 
$ \rm \log(\frac ZX) \simeq [Fe/H] + \log(\frac ZX)_\odot $. 
We use the solar mixture of Grevesse \& Noels\,(\cite{gn93}):
$\left(\frac Z X\right)_\odot = 0.0245 $.

The evolution tracks are initialized at the Pre-Main Sequence stage.
Note that the age is counted from the ZAMS. In CESAM, the ZAMS is defined
  as the stage of the end of the Pre-Main Sequence where the gravitational energy 
  release is equal to the nuclear one. 
We have computed models with and without microscopic diffusion of chemical species.


To fit observational data (effective temperature $ \rm T_{eff}$~, 
  luminosity $ \rm L $ and surface metallicity $ \rm [Z/X]_{surf} $)
  with corresponding results of various computations,
  we adjust the main stellar modeling parameters:  mass, 
  age and metallicity.
In figures (Figs \ref{fig:Eri_zoom}, \ref{fig:Hya_zoom}, 
  \ref{fig:Boo_zoom} and \ref{fig:Boo-diff_zoom}) 
  representing the zoom of HR diagram, the (rectangular) error boxes are
  derived from the values and accuracies of the stellar parameters quoted
  in Table \ref{tableau}. 
The present (new) values of radii, presented in this paper, select sub-areas
  in these error boxes and hence

the new measures of diameters are used to discriminate our models (see Table \ref{tableau}). 
Our best model is designed as the one which satisfies first the luminosity and radius constraint
   and second the effective temperature constraint. 
On the zooms of the HR diagrams (see Figures \ref{fig:Eri_zoom}, \ref{fig:Hya_zoom}, 
   \ref{fig:Boo_zoom} and \ref{fig:Boo-diff_zoom}), the measured radius
   and its confidence interval appear as diagonal lines. 
We notice that the addition of the radius measurement reduces significantly 
   the uncertainty domain, and in some cases tightens the allowed range for 
   ages by a factor three (see below).
We have computed models that include overshooting of the convective
  core (radius $ R_{co}$) over the distance 
  $ O_{\rm v} = A_{ov} \min(H_{\rm p}, R_{\rm co})$ where $R_{\rm co}$ is the core radius, 
  following the prescriptions of Schaller et al. (\cite{schaller}). 

\subsection{$\delta$ Eri}
\label{delta-Eri}

First, we adopt an initial helium content similar to the Sun 
  $ \rm Y_{ini} = 0.28 $ and $ \rm [Z/X]_{ini} = 0.148 $,
  both stars having similar ages and abundances (this will be confirmed 
  hereafter). 

\begin{figure}[t]
\centering
\includegraphics[angle=-90, width=8.8cm]{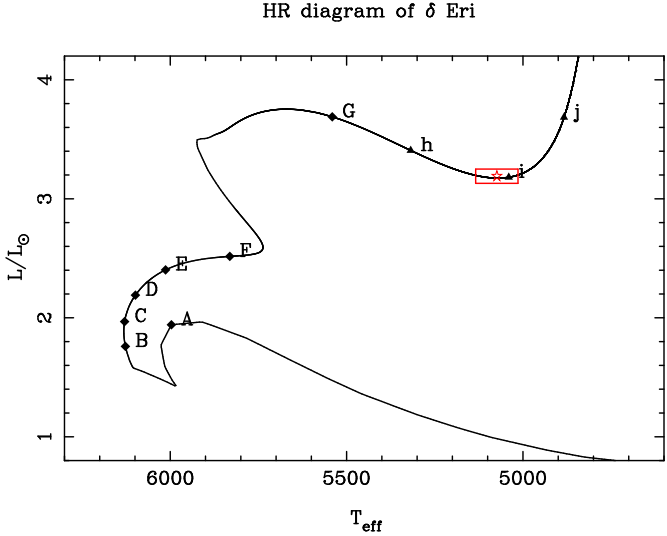}
\caption{ \label{fig:Eri_hr} Evolutionary tracks in the H-R diagram for $\delta$ Eri
  from label 'A' (0. Myr) to label  'G' (6000. Myr), shown by upper case letters and
  squares with time steps of 1000. Myr; from label 'h' (6100. Myr) to label 'j' (6300. Myr),
  shown by lower case letters and triangles with time steps of 100. Myr. }
\end{figure}

\begin{figure}[t]
\centering
\includegraphics[angle=-90, width=8.8cm]{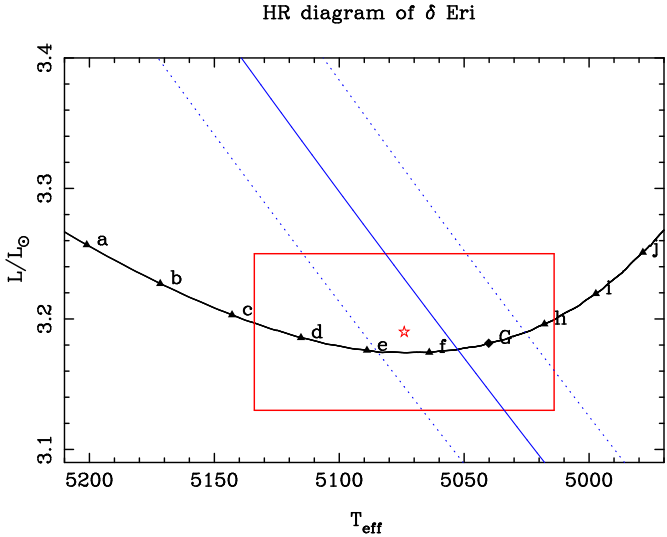}
\caption{ \label{fig:Eri_zoom} Zoom of the evolutionary tracks in the H-R diagram for $\delta$ Eri
   from label 'a' (6140. Myr) to label 'j' (6230. Myr), shown by lower case letters and
   triangles with time steps of 10. Myr (except label 'G' at 6200. Myr shown by an upper case letter and 
   a square). Our best model is close to label 'f' at 6194. Myr (see table \ref{tableau}). }
\end{figure}

Then, with mass and metallicity as free parameters, we have computed
  a grid of evolutionary tracks in order to reproduce observational 
  data. 
Our best model without diffusion and without overshooting gives:
  $ \rm M = 1.215 \, M_\odot $ and an age (from the ZAMS) of $ \rm 6196. \, Myr$. 
Our best model with diffusion and an overshooting value of $ A_{ov} = 0.15 $ 
  in agreement with the results of Ribas et al. (\cite{ribas00}) gives:
  $ \rm M = 1.215 \, M_\odot $, an age (from the ZAMS) of $ \rm 6194. \, Myr $ 
  and a diameter of $ \rm D = 2.328 \, D_\odot $. 
See Figures \ref{fig:Eri_hr} and \ref{fig:Eri_zoom}. 
 
%
%
%


The mean large frequency splitting found for our best model is $ \rm 45.27 \, \mu Hz$. 
This result is in agreeement within two per cent with the value of $ \rm 43.8 \, \mu Hz$
   of the mean large frequency splitting reported by Carrier et al. (\cite{carrier03}).

\subsection{$\xi$ Hya}
\label{ksi-Hya}

\begin{figure}[t]
\centering
\includegraphics[angle=-90, width=8.8cm]{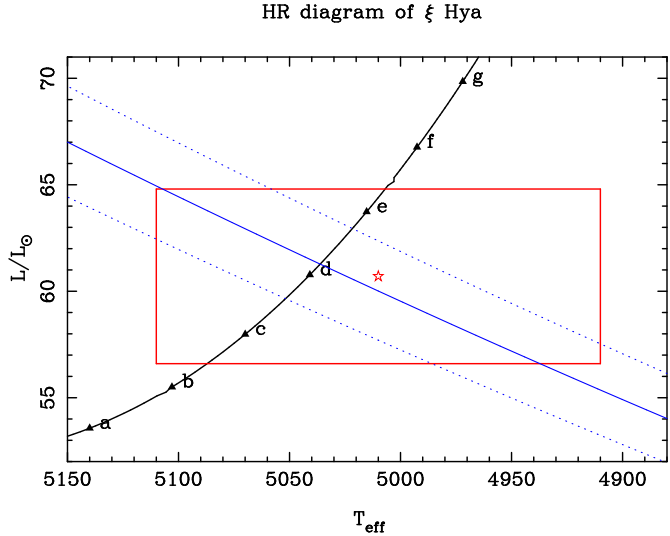}
\caption{ \label{fig:Hya_zoom} Zoom of the evolutionary tracks in the H-R diagram for $\xi$ Hya. 
   from label 'a' (509.2 Myr) to 'g' (509.8 Myr), shown by lower case letters and triangles with
   time steps of 0.1 Myr. Our best model is close to label 'd' at 509.5 Myr (see table \ref{tableau}).
}
\end{figure}

We have computed a grid of evolutionary tracks (with and without diffusion)
  in order to reproduce observational data. 
Hence, we derived the following parameters: 
  $ \rm M = 2.65 \, M_\odot $ , $ \rm Y_{ini} = 0.275 $ and $ \rm [Z/X]_{ini} \equiv  0.0 $. 
Our best model with diffusion and an overshooting value of $ A_{ov} = 0.20 $ 
  in agreement with the results of Ribas et al. (2000) gives us
  an age (from the ZAMS) of $ \rm 509.5 \, Myr $ and a diameter of $ \rm D = 10.3 \, D_\odot $. 
To improve the modeling, a better precision of the diameter is required 
  as it is the case for the two other stars discussed in this paper, 
  for which the accuracy is better by an order of magnitude. 
See Figures \ref{fig:Hya_hr} and \ref{fig:Hya_zoom}.

\begin{figure}[t]
\centering
\includegraphics[angle=-90, width=8.8cm]{ksi-Hya_hr_sp.ps}
\caption{ \label{fig:Hya_hr} Evolutionary tracks in the H-R diagram for $\xi$ Hya 
   from label 'A' (0. Myr) to  'F' (500. Myr), shown by upper case letters and
   squares with time steps of 100. Myr; from label 'g' (502. Myr) to 'p' (511. Myr),
   show by lower case letters and triangles with time steps of 1. Myr. 
}
\end{figure}
%

Solar-like oscillations of that star were discovered by Frandsen et al. (\cite{frandsen02}) 
  with a mean spacing of $ \rm 7.1 \, \mu Hz$  
  see also Teixeira et al. (\cite{teixeira03})).
  From our model, we computed a value of 
  $ \rm 7.2 \, \mu Hz $ similar to the theoretical value presented by 
  Frandsen et al.  or  Teixeira et al. . 

\subsection{$\eta$ Boo}
\label{eta-Boo}

\begin{figure}[t]
\centering
\includegraphics[angle=-90, width=8.8cm]{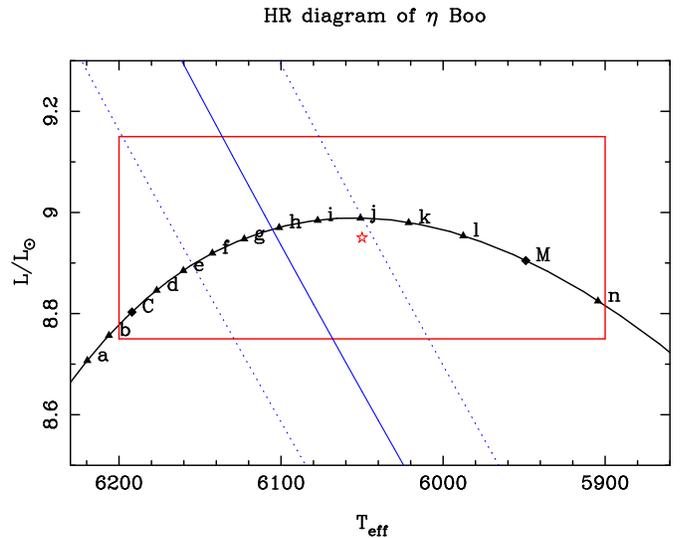}
\caption{ \label{fig:Boo_zoom} Zoom of the evolutionary tracks in the H-R diagram for $\eta$ Boo (model without            diffusion) from label 'a' (2280. Myr) to 'n' (2410. Myr), shown by lower case letters and triangles with
  time steps of 10. Myr (except labels 'C' at 2300. Myr and label 'M' a 2400. Myr shown by an upper case 
  letters and squares). Our best model is close to label 'h' at 2350. Myr (see table \ref{tableau}).
}
\end{figure}

\begin{figure}[t]
\centering
\includegraphics[angle=-90, width=8.8cm]{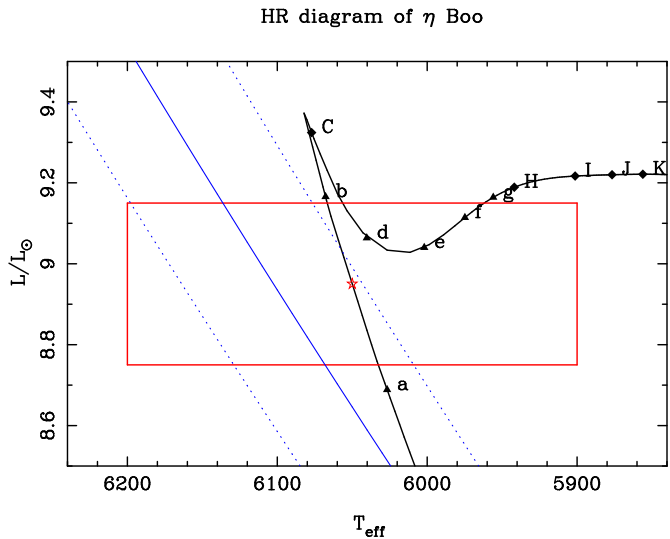}
\caption{ \label{fig:Boo-diff_zoom} Zoom of the evolutionary tracks in the H-R diagram for $\eta$ Boo 
  (model with diffusion) from label 'a' (2738. Myr) to label 'g' (2744. Myr), shown by lower case letters
  and triangles with time steps of 1. Myr (except label 'C' at 2740. Myr shown by a square); from 
  label 'H' (2745. Myr) to label 'K' (2760. Myr), shown by upper case letters and squares with time
  steps of 5. Myr. Our best model is between label 'a' (at 2738. Myr) and label 'b' (at 2739. Myr)
  (see table \ref{tableau}).}
\end{figure}

Concerning the values of $ \rm T_{eff} $ and its corresponding uncertainty, we
   have chosen conservative values based upon various 
   determinations:
   Feltzing \& Gonzales (\cite{feltzing01}) gives $ \rm T_{eff} = 6000. \pm 100. \, K $ whereas 
   Cayrel de Strobel (\cite{giusa01}) gives a range between 5943. et 6219. K . 
We notice that DiMauro et al. adopt $ \rm T_{eff} = 6028. \pm 45. \, K $ but
   in our study, we take advantage of the constraint given by the new diameter value which
   reduces the uncertainty as shown on Figures \ref{fig:Boo_zoom} or \ref{fig:Boo-diff_zoom}. 
   
In a first attempt to characterize this star, DiMauro et al. (\cite{dimauro03}) propose 
  to limit the range of mass between  $ \rm 1.64 \, M_\odot $ and $ \rm 1.75 \, M_\odot $. 
Recently, Guenther (\cite{guenther04}) adopted in his conclusion a mass of $ \rm 1.706 \, M_\odot $
  with an initial chemical composition: $ \rm X_{ini} = 0.71 $ , $ \rm Y_{ini} = 0.25 $ and $ \rm Z_{ini} = 0.04 $. 
In the present study, we have computed a grid of models and it appears that the
  best fitting parameters are: $ \rm M =  1.70 \, M_\odot $ with an initial chemical composition:
  $ \rm X_{ini} = 0.70 $ , $ \rm Y_{ini} = 0.26 $ and $ \rm Z_{ini} = 0.04 $. 
A first set of models have been computed with the simplest available but reliable 
  physics (and therefore without diffusion, as probably done by the previous cited authors).
A second set of models have also been computed with improved physics.
Thus, we include convective overshooting (with $ A_{ov} = 0.15 $, see previous discussion),
  diffusion and radiative diffusivity (see Morel \& Th\'evenin \cite{morel02}) which
  controls diffusion of chemical elements in intermediate mass stars. 
The two sets of results give evidently similar results except for the ages: the age of the
  best model with diffusion (2738.5 Myr) is larger than the age of the best model without diffusion (2355.0 Myr).  

\begin{figure}[t]
\centering
\includegraphics[angle=-90, width=8.8cm]{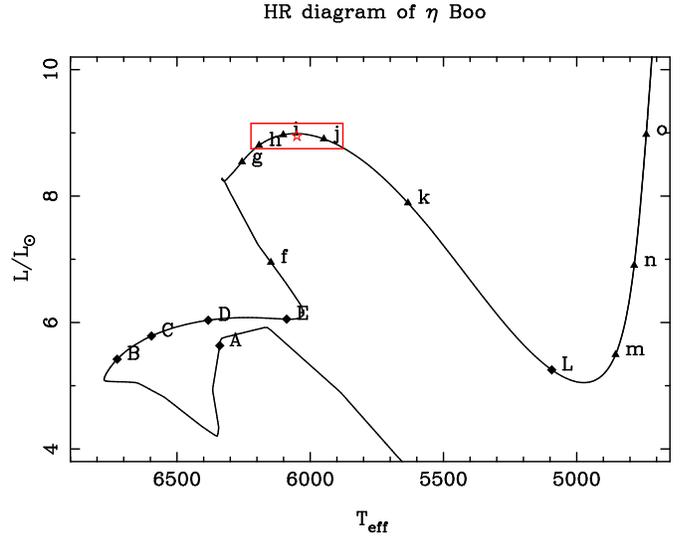}
\caption{ \label{fig:Boo_hr} Evolutionary tracks in the H-R diagram for $\eta$ Boo (model without diffusion)
   from label 'A' (0. Myr) to  'E' (2000. Myr), shown by upper case letters and squares with time steps of
   500. Myr; from label 'f' (2200. Myr) to 'o' (2650. Myr), shown by lower case letters and triangles with
   time steps of 50. Myr (except label 'L' at 2500. Myr shown by an upper case letter and a square). 
}
\end{figure}

\begin{figure}[t]
\centering
\includegraphics[angle=-90, width=8.8cm]{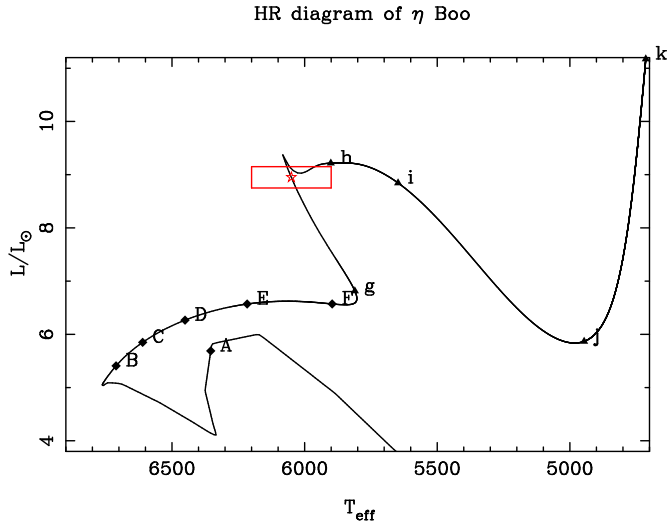}
\caption{ \label{fig:Boo-diff_hr} Evolutionary tracks in the H-R diagram for $\eta$ Boo (model with diffusion)
  from label 'A' (0. Myr) to label 'F' (2500. Myr), shown by upper case letters and squares with a
  time step of 500. Myr; label 'g' at 2700. Myr shown by a triangle; from label 'h' (2750. Myr) to
  label 'k' (2900. Myr), shown by lower case letters and triangles with a time step of 50. Myr.
}
\end{figure}

As shown, for example, on Figure \ref{fig:Boo_zoom}, without the constraint given by the
   diameter, the age would be ranging from 2295. Myr (between label 'b' and label 'c')   
   to 2410. Myr (close to label 'n'), with a derived uncertainty of 115. Myr.   
For a given set of input physics, the constraint on diameter reduces the uncertainty
   on the age by about a factor three : the age would be ranging from 2323. Myr (close to label 'e') 
   to 2370. Myr (close to label 'j'), corresponding to a (reduced) uncertainty of 47. Myr 
   (Figures \ref{fig:Boo_hr}, \ref{fig:Boo_zoom}, \ref{fig:Boo-diff_hr} and \ref{fig:Boo-diff_zoom}). 
Note that our model for $\eta$ Boo with diffusion (Figures \ref{fig:Boo-diff_hr} and \ref{fig:Boo-diff_zoom})
   has the star in a very short-lived phase of evolution (which is, of course, possible but with a small, 
   but non zero, probability). 
%
%
%
%
\section{Concluding remarks}
We have measured with the instrument VLTI/VINCI the angular diameters of three subgiant and giant 
  stars and used them as an additive constraint to the spectro-photometric and asteroseismic 
  ones to perform a study of their evolutionary status. 

Owing the position of the three stars in the HR diagram, we can notice that 
   the determination of the modeling parameters, in particulary the age, is 
   very sensitive to the input physics, due to the rapidity of the stellar 
   evolution compared to the size of the error boxes. 

With our input physics and observational constraints, 
  $\delta$ Eri is a star at the end of the subgiant phase 
  ($ \rm M = 1.215 \, M_\odot $) with an age of 6.2 Gyr.
  We attempt without success to detect a close companion forcing us to conclude
  that the classification of $\delta$ Eri as an RS CVn star is doubtful.

$\xi$ Hya has been constrained with success with a model adopting a mass of $ \rm 2.65 \, M_\odot $ 
  and an age of 510. Myr. 
 
$\eta$ Boo is a subgiant slightly more evolved than Procyon with a similar
   age of 2.7 Gyr. 
With a mass of at $ \rm M = 1.7 \, M_\odot $ 
   (similar to the mass adopted by Di Mauro et al. (\cite{dimauro03})), 
   we were able to reproduce the VLTI/VINCI radius. 
We notice that because of the short evolutionary time scales of a model 
   crossing rather large error boxes, the results of the models -- in
   particular the age -- are very sensitive to the input physics (for
   instance, the core mixing.
Some progress on the asteroseismic observations are now required to better 
   constrain the evolution state of giant stars for which the frequency spacings 
   (Bouchy \& Carrier \cite{bouchy03}, Bedding \& Kjeldsen \cite{bedding03}) 
   are still relatively imprecise. 
The improvement of the angular diameter estimations in the future will further 
   tighten the uncertainty domain on the HR diagram, especially as detailed 
   modeling of the atmosphere will be required. 
This improvement will naturally require a higher precision on the parallax 
   value to derive the linear diameters.

\begin{acknowledgements}
The VINCI public commissioning data reported in this paper
   has been retrieved from the ESO/ST-ECF Archive.
The VINCI pipeline includes the wavelets processing technique,
  developed by D. S\'egransan (Obs. de Gen\`eve).
No VLTI observation would have been possible without the efforts of the
  ESO VLTI team, to whom we are grateful.
This work has been performed using the computing facilities provided 
  by the program Simulations Interactives et Visualisation en Astronomie 
  et M\'ecanique (SIVAM) at the computer center of the Observatoire de la
  C\^ote d'Azur. 
This research has made use of the Simbad database operated at CDS, Strasbourg, France. 
We thank the referee, T. R. Bedding, for his suggested improvements of this paper.
\end{acknowledgements}
{}
\end{document}